\documentclass[12pt,a4paper]{article}
\usepackage{graphicx,color}
\usepackage{amsmath}
\usepackage{amssymb}
\pdfoutput=1
\usepackage{multirow}
\usepackage{url}
\usepackage{subfigure}
\usepackage{authblk}
\usepackage{fullpage}
\usepackage{lscape}

\begin{document}
\vspace*{-0.5in}
\thispagestyle{empty}

\vspace{0.5in}
\begin{center}
\begin{LARGE}
{\bf Status Report (22th J-PARC PAC):\\
Searching for a Sterile Neutrino at J-PARC MLF (E56, JSNS$^2$)\\}
\vspace{5mm}
\end{LARGE}
\begin{large}
\today\\
\end{large}
\vspace{5mm}
{\large
M.~Harada, S.~Hasegawa, Y.~Kasugai, S.~Meigo,  K.~Sakai, \\
S.~Sakamoto, K.~Suzuya \\
{\it JAEA, Tokai, JAPAN}\\
\vspace{2.0mm}
T.~Maruyama\footnote{{Spokesperson: 
(takasumi.maruyama@kek.jp) }}, 
S.~Monjushiro, K.~Nishikawa, M.~Taira\\
{\it KEK, Tsukuba, JAPAN}\\
\vspace{2.0mm}
S.~Iwata, T.~Kawasaki \\
{\it Department of Physics, Kitasato University, JAPAN}\\
\vspace{2.0mm}
M.~Niiyama\\
{\it Department of Physics, Kyoto University, JAPAN}\\
\vspace{2.0mm}
S.~Ajimura, T.~Hiraiwa, T.~Nakano, M.~Nomachi, T.~Shima, Y.~Sugaya\\
{\it RCNP, Osaka University, JAPAN}\\
\vspace{2.0mm}
T.~J.~C.~Bezerra, E.~Chauveau, H.~Furuta, Y.~Hino, F.~Suekane\\
{\it Research Center for Neutrino Science, Tohoku University, JAPAN}\\
\vspace{2.0mm}
I.~Stancu\\
{\it University of Alabama, Tuscaloosa, AL 35487, USA}\\
\vspace{2.0mm}
M.~Yeh\\
{\it Brookhaven National Laboratory, Upton, NY 11973-5000, USA}\\
\vspace{2.0mm}
W.~Toki\\
{\it Colorado State University, Fort Collins, Colorado, USA}\\
\vspace{2.0mm}
H.~Ray\\
{\it University of Florida, Gainesville, FL 32611, USA}\\
\vspace{2.0mm}
G.~T.~Garvey, C.~Mauger, W.~C.~Louis, G.~B.~Mills, R.~Van~de~Water\\
{\it Los Alamos National Laboratory, Los Alamos, NM 87545, USA}\\
\vspace{2.0mm}
E.~Iwai, J.~Jordan, J.~Spitz\\
{\it University of Michigan, Ann Arbor, MI 48109, USA}
}
\end{center}
\renewcommand{\baselinestretch}{2}
\large
\normalsize

\setlength{\baselineskip}{5mm}
\setlength{\intextsep}{5mm}

\renewcommand{\arraystretch}{0.5}

\newpage

\tableofcontents
\vspace*{0.5in}
\setcounter{figure}{0}
\setcounter{table}{0}
\indent

\section{Introduction}
\indent

The JSNS$^2$ (J-PARC E56) experiment aims to search for a sterile neutrino at the
J-PARC Materials and Life Sciences Experimental Facility (MLF).
After the submission of a proposal~\cite{CITE:PROPOSAL} to the J-PARC PAC,
Stage-1 approval was granted to the JSNS$^2$ experiment on April 2015.
This approval followed a series of background measurements which were performed
in 2014~\cite{CITE:SR_14NOV, CITE:SR_14NOV2}.

Recently, funding (the grant-in-aid for scientific research (S)) in Japan for building one 25~ton fiducial volume detector module was approved for the experiment. Therefore, we aim to start the experiment with one detector in JFY2018-2019. 
We are now working to produce precise cost estimates and schedule for construction, noting that most of the detector components
can be produced within one year from the date of order. This will be reported
at the next PAC meeting.  

In parallel to the detector construction schedule, JSNS$^2$ will submit a
Technical Design report (TDR) to obtain the Stage-2 approval from the J-PARC
PAC.
The recent progress of the R$\&$D efforts towards this TDR
are shown in this report. In particular, the R$\&$D status of the liquid
scintillator, cosmic ray veto system, and software are shown. 

We have performed a test-experiment using 1.6~L of liquid scintillator at the 3rd
floor of the MLF building in order to determine the identities of non-neutrino background particles coming to this detector location during the proton bunch. This is the so-called
``MLF 2015AU0001'' experiment. We briefly show preliminary results from this test-experiment.

\section{\setlength{\baselineskip}{4mm} Timescale and Cost estimation for Detector Construction}
\indent

The following detector elements are considered for developing the JSNS$^2$ experiment.
\begin{itemize}
\item A stainless tank and an acrylic tank
\item PMTs
\item Liquid scintillator
\item Veto system
\item Electronics
\item Slow components such as HV, slow monitors, and safety equipment.
\end{itemize}  

The stress calculation for the stainless tank was done and submitted to
the PAC already.
%~\cite{CITE:SR_14NOV}.
We use 8'' PMTs instead of 10'' PMTs in order to 
to optimize for the dynamic range of the PMT and electronics, however the oil
(liquid scintillator) protection mechanism for the breeder chain is the same
as for Double-Chooz and RENO~\cite{CITE:PMT}.
The tapered breeder system to expand the non-saturation region between
the input light yield and the output signal was tested and discussed in the
status report~\cite{CITE:21PAC}. 
A brief discussion of the cosmic ray veto system is presented later in this status
report. 

The other items involve finalizing the cost estimation and production time
scale. The TDR will describe not only the design of these sub-systems, but also
the relevant contracts and time-scale.

\section{Status of Studies for the TDR}

\subsection{Liquid Scintillator}
\indent

\begin{figure}
 \centering
 \includegraphics[width=0.7\textwidth]{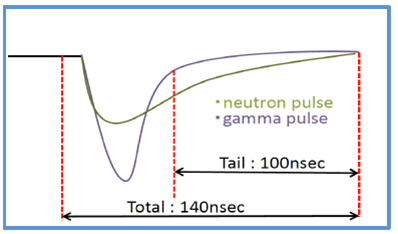}
 \caption{\setlength{\baselineskip}{4mm}
   The PSD variable concept. Neutron background events have wider pulse
   shape than that of gamma / positron events. The PSD variable is defined
   as tailQ/totalQ (where Q is charge).
 }
 \label{fig:PSDcon}
\end{figure}

One of the most important goals of the JSNS$^2$ R$\&$D is to produce a detector that is is able to efficiently and purely discriminate fast neutron background events from signal events. One technique for distinguishing these signals is called Pulse Shape Discrimination (PSD), which takes advantage of the different timing characteristics of signal and background.
The definition of the PSD variable is shown in the
Fig.~\ref{fig:PSDcon}. As seen in the conceptual plot, the neutron
background events have wider waveforms than that of signal events.

The other technique for distinguishing between signal and background is to use Cherenkov photons. An oscillated electron antineutrino IBD event is characterized by a final state positron, which emits
Cherenkov light. However, neutron-induced proton events do not emit Cherenkov light at these low energies.

The JSNS$^2$ progress on liquid scintillator R$\&$D has been shown elsewhere~\cite{CITE:20PAC,CITE:21PAC}, 
therefore we only show the most relevant recent results.

%***************
\subsubsection{\setlength{\baselineskip}{4mm}
Effect of Noise on PSD capability}
\indent

\begin{figure}[htpb]
\centering
\includegraphics[width=0.8 \textwidth]{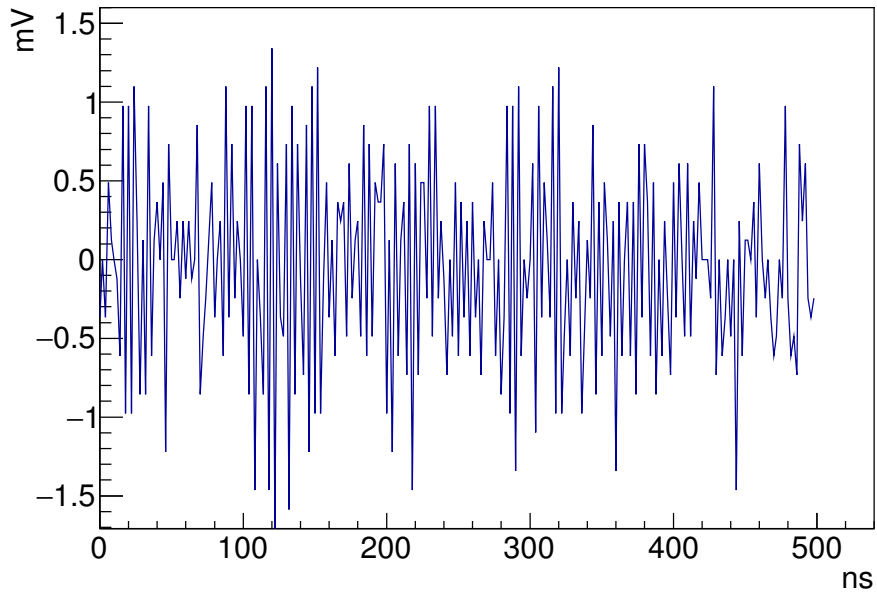}
\caption{\setlength{\baselineskip}{4mm}
 An example of the measured noise waveform with an 8 inche PMT at Tohoku University. 
}
\label{FIG:NOISEDATA}
\end{figure}

\begin{figure}[htpb]
\centering
\includegraphics[width=1.0 \textwidth]{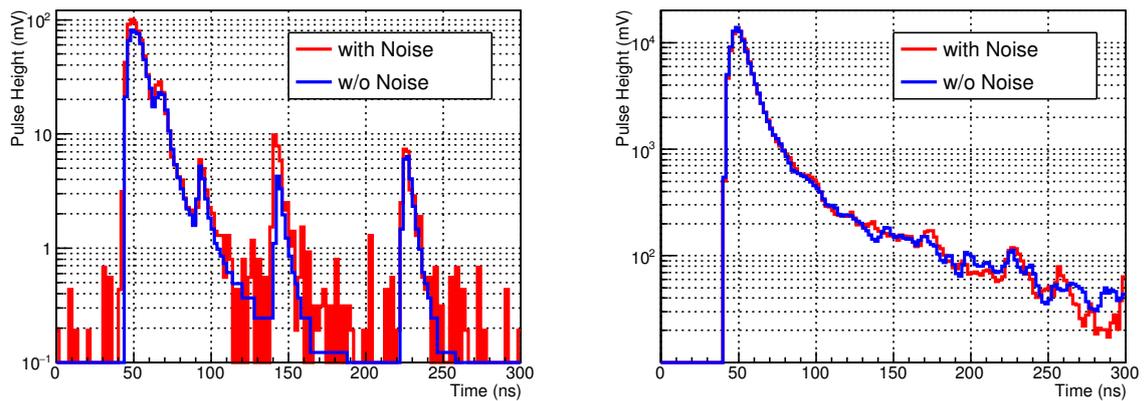}
\caption{\setlength{\baselineskip}{4mm}
An example of a waveform from one PMT (left plot, 51~p.e.s) and the sum of the waveforms from all PMTs (right plot, 9200~p.e.s, 45MeV) in the neutrino MC samples. Blue and red lines show cases of with and without the noise data, respectively.
}
\label{FIG:NOISEWF}
\end{figure}

In a previous status report~\cite{CITE:21PAC}, the estimated PSD capability of a full JSNS$^{2}$ detector with Daya Bay type Gd-loaded liquid scintillator(DBLS) was shown. This work was based on a MC study with neutrinos (signal) and cosmic-induced fast neutron samples after applying the neutrino selection criteria. However, the noise effect of the PMTs was not considered in the study at that time.
Scintillation light in the JSNS$^{2}$ detector is viewed by a few hundred PMTs, and the PSD capability is evaluated with sum of the waveforms of all the PMTs. Compared with detecting the light using one PMT, the noise can affect the PSD capability considerably. Therefore, it is important for a more realistic PSD study in consideration of PMT noise. 
The PSD capability in the previous status report was recalculated implementing measured noise data with an 8 inche PMT (Hamamatsu R5912). Figure~\ref{FIG:NOISEDATA} shows an example of the measured noise waveform at Tohoku University. 
Figure~\ref{FIG:NOISEWF} shows an example of waveforms from one PMT(the left figure) and the sum of the waveforms of all PMTs(the right figure) for the neutrino MC samples before and after including the noise data. 

\begin{figure}[htpb]
\centering
\includegraphics[width=0.8 \textwidth]{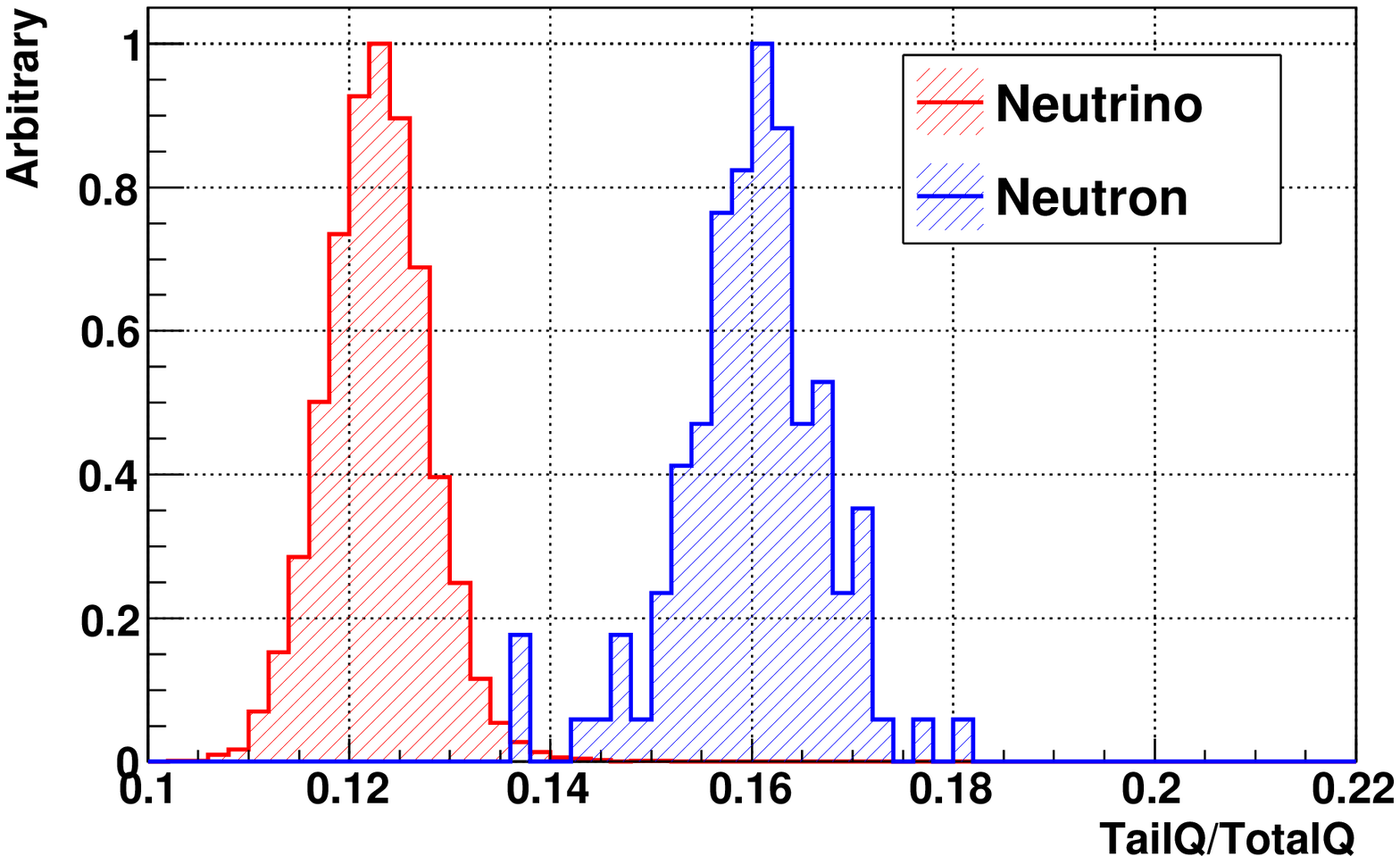}
\caption{\setlength{\baselineskip}{4mm}
PSD distributions of the neutrino MC samples (red line) and the cosmic-induced fast neutrons (blue line, recoiled protons) in the previous status report. 
}
\label{FIG:PSD}
\end{figure}

\begin{figure}[htpb]
\centering
\includegraphics[width=0.8 \textwidth]{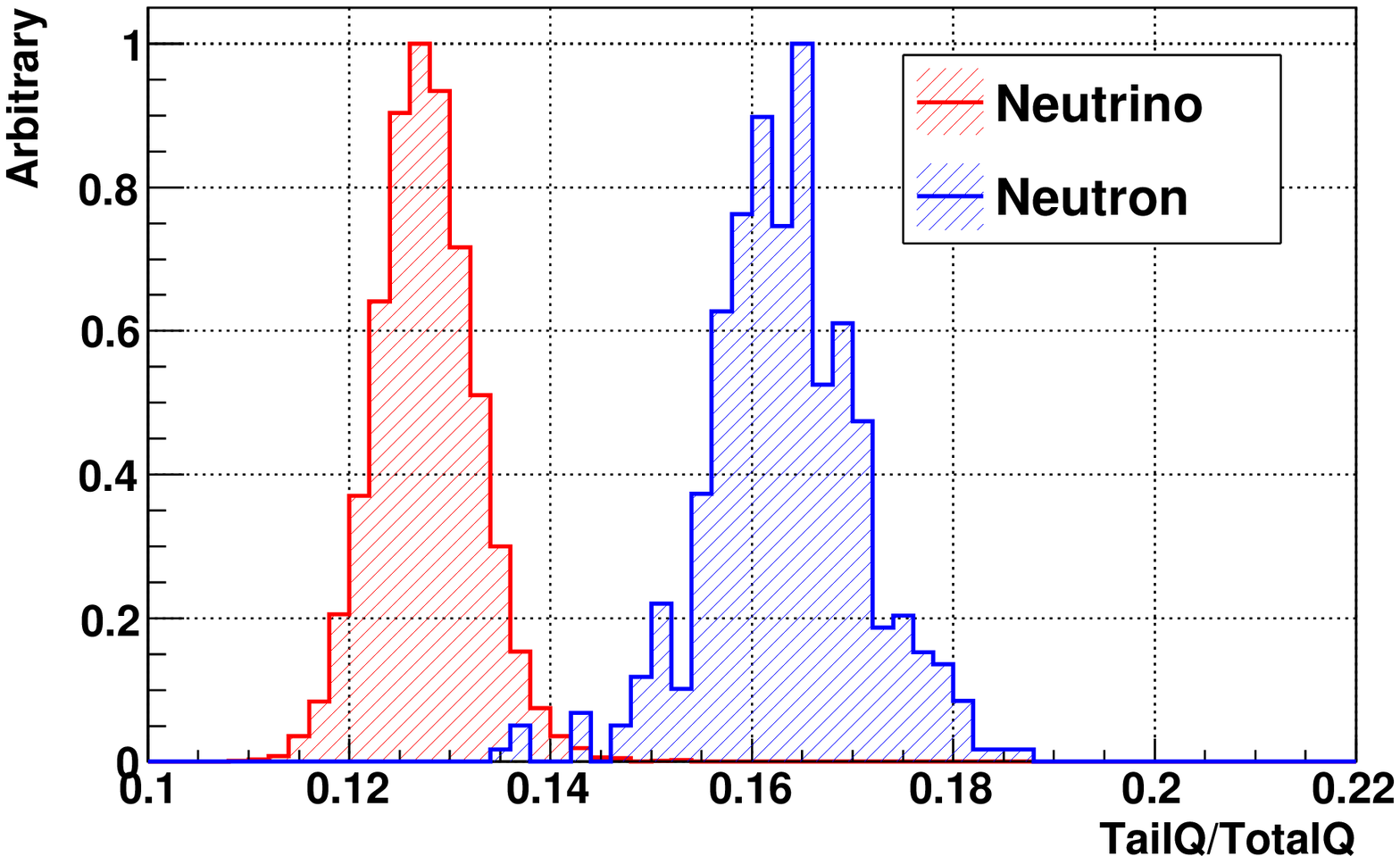}
\caption{\setlength{\baselineskip}{4mm}
PSD distributions of the neutrino MC samples and the cosmic-induced fast neutrons (recoiled protons) after including the normal noise. 
}
\label{FIG:PSDNORMALNOISE}
\end{figure}

\begin{figure}[htpb]
\centering
\includegraphics[width=0.8 \textwidth]{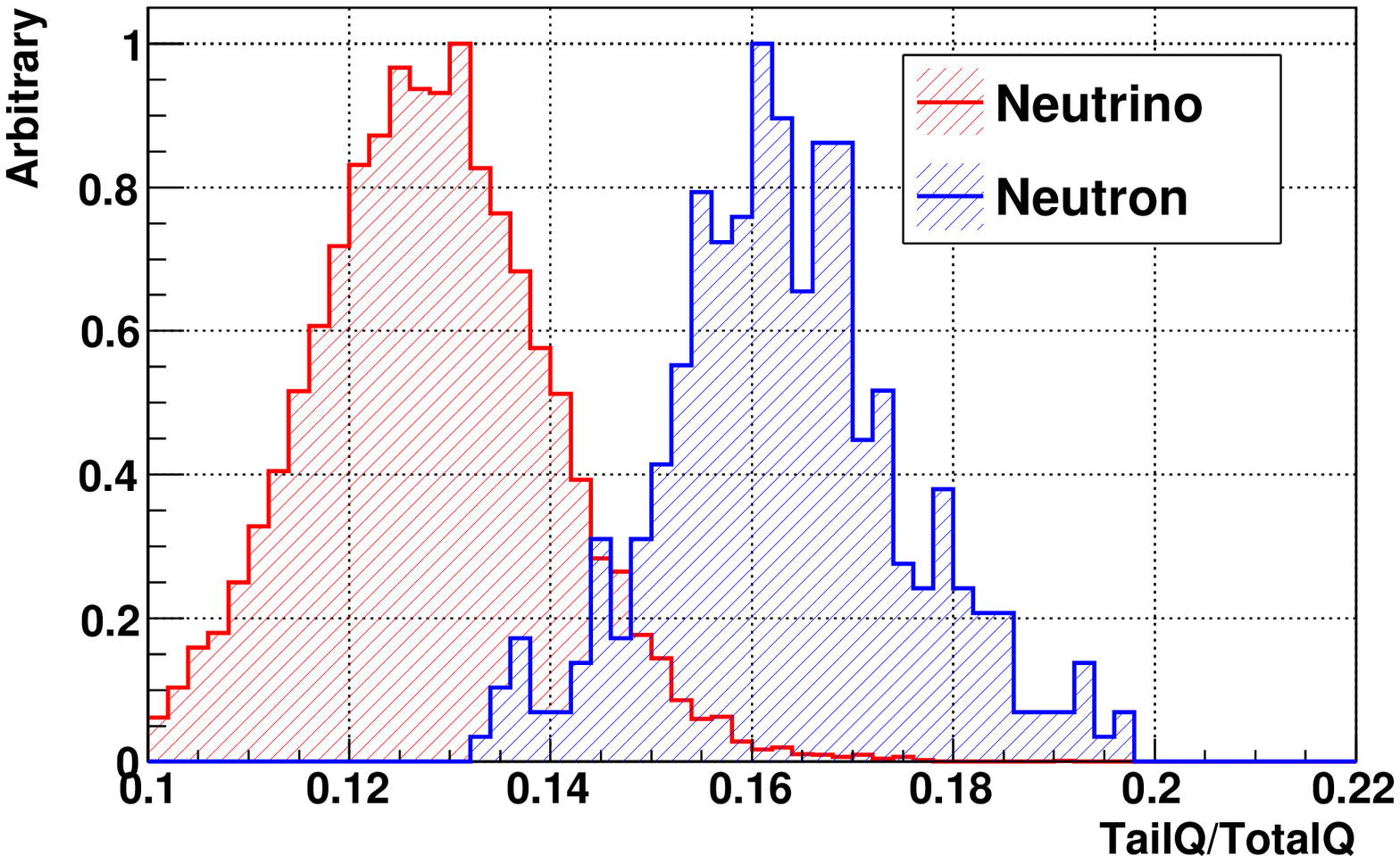}
\caption{\setlength{\baselineskip}{4mm}
PSD distributions of the neutrino MC samples and the cosmic-induced fast neutrons (recoiled protons) after including the coherent noise.
}
\label{FIG:PSDCOHE}
\end{figure}

Two noise cases were considered, independent noise for each PMT (normal noise) and same noise for all PMTs (coherent noise) for the extreme case.
Figures ~\ref{FIG:PSD}, \ref{FIG:PSDNORMALNOISE} and \ref{FIG:PSDCOHE} show the PSD distributions of the previous study, the normal noise case, and the coherent noise case, respectively. 

The PSD distribution of the normal noise case is not changed so much comparing the previous result. However, the PSD distribution of the coherent noise indicates that the coherent noise affects the PSD capability more than the normal noise case. The coherent noise could be made inside electronics for data taking. It is necessary that the coherent noise is considered in the development of the electronics for JSNS$^2$.
More quantitative analysis of the noise effect will be done in the near future.

%***************

\subsubsection{\setlength{\baselineskip}{4mm} New PSD Results using DIN-based Liquid Scintillator}
\indent

In this subsection, we show new PSD results using DIN-based liquid
scintillator.

\begin{figure}[h]
\centering
\includegraphics[width=.5\textwidth]{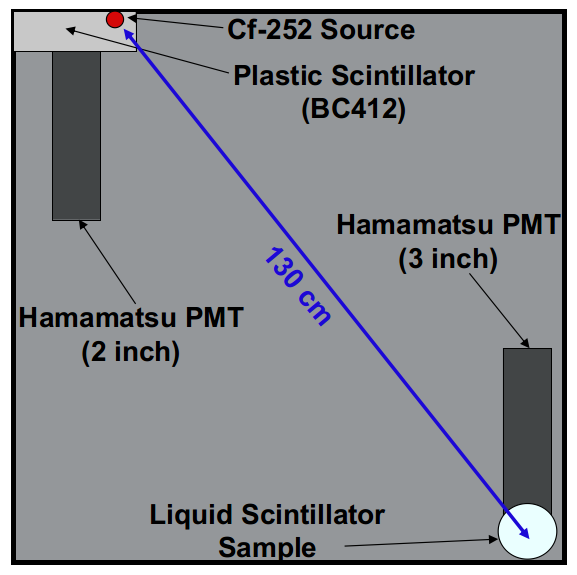}
\caption{\setlength{\baselineskip}{4mm} Schematic diagram of the dark box used for our scintillator tests. PMTs and scintillators not shown to scale.}
\label{TestStand}
\end{figure}

A schematic diagram of the test stand used to evaluate the scintillator mixtures is shown in Fig.~\ref{TestStand}. A Cf-252 source was placed near a piece of plastic scintillator with a PMT (Hamamatsu H6410) viewing it. Another PMT (Hamamatsu H6559) was placed approximately 130 cm away looking at the liquid scintillator sample being tested. In these tests, we used an LAB-based scintillator cocktail (LAB + 3.0 g/L PPO + 15 mg/L bis-MSB) and a DIN-based scintillator cocktail (DIN + 3.0 g/L PPO + 15 mg/L bis-MSB) provided by BNL. Both scintillator cocktails were bubbled with nitrogen for 1 hour to remove impurities that quench scintillation light. We looked for coincident hits in both the plastic and liquid scintillators to identify Cf decays. Signals were read out using a 14-bit CAEN DT5730 digitizer at 500 MS/s. 

\begin{figure}[h] 
\centering
\includegraphics[width=.45\textwidth]{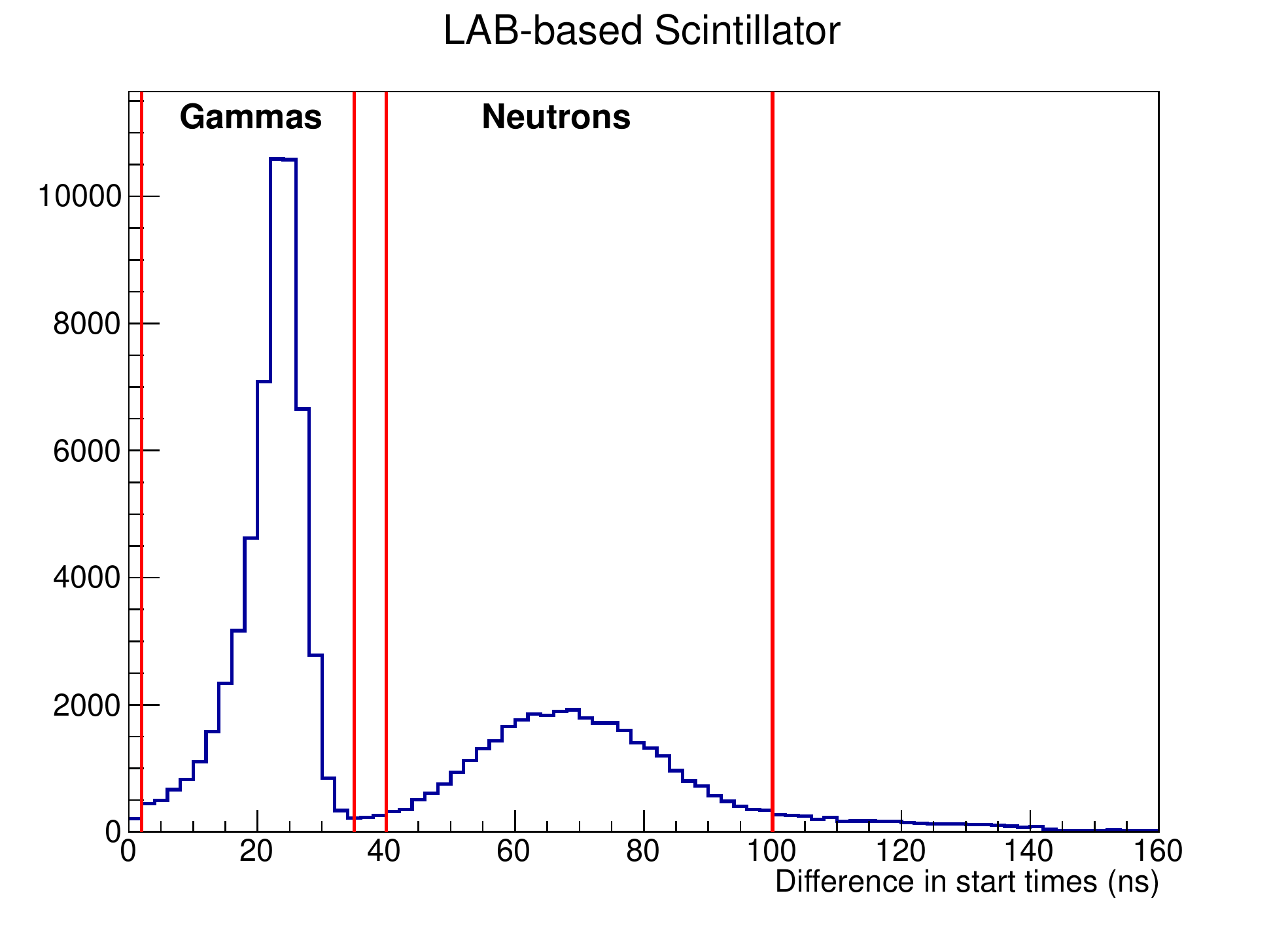}
\includegraphics[width=.45\textwidth]{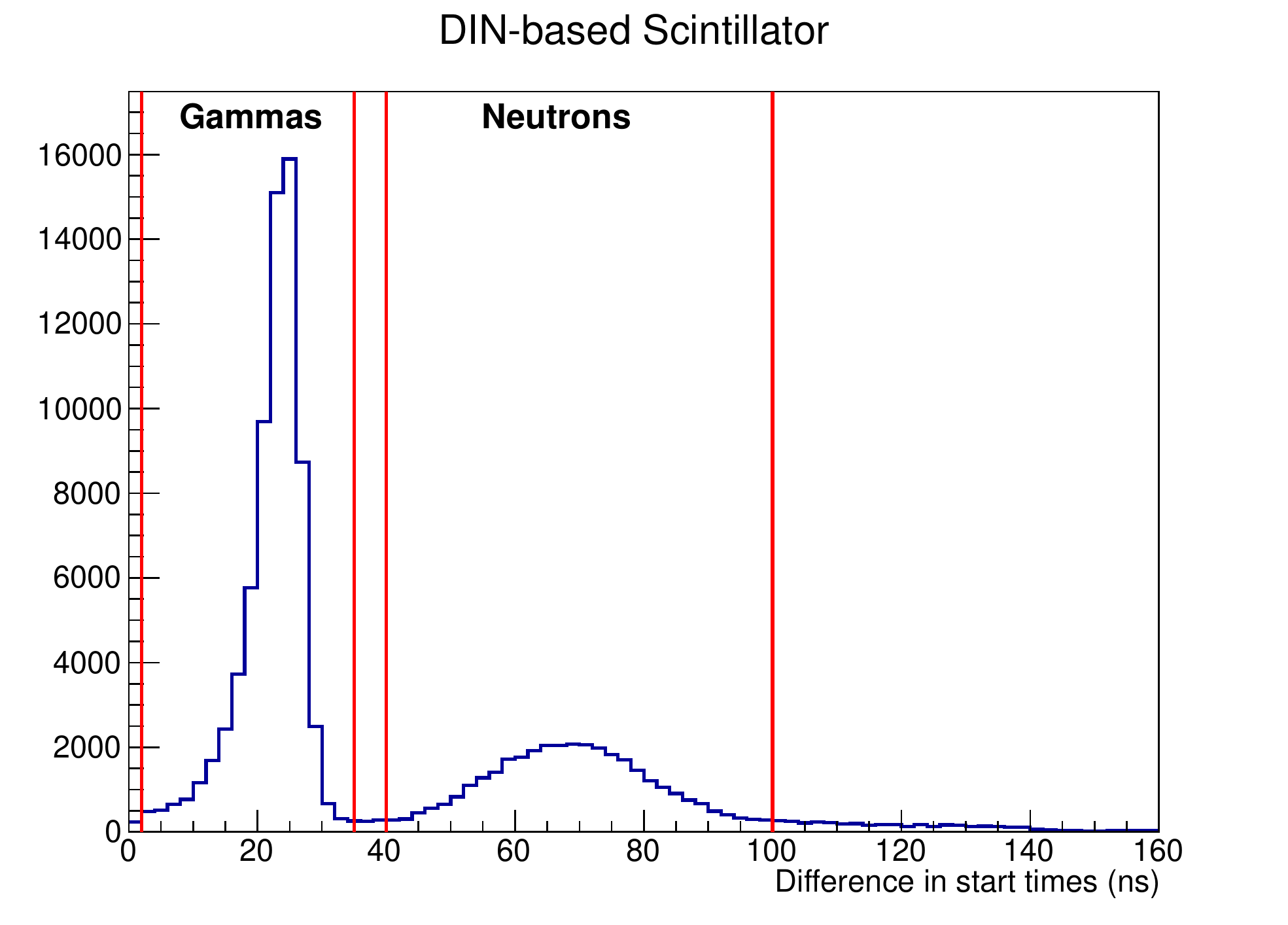}
\caption{\setlength{\baselineskip}{4mm} Time of flight for events in the LAB-based scintillator (left) and the DIN-based scintillator (right). The population at low time of flight values is gammas and the population with longer time of flight values contains neutrons. The red lines mark the boundaries of the regions used to classify events as gammas and neutrons.}
\label{TOFDists}
\end{figure}

To determine if an event in the liquid scintillator was a neutron or a gamma, we used time of flight (TOF) information. By measuring the time difference between the triggers in the liquid and plastic scintillators it was possible to tell if an event was a gamma (short TOF) or a neutron (long TOF). Histograms of the time of flight distributions for two different scintillator cocktails are shown in Fig.~\ref{TOFDists}.

\begin{figure}[h]
\centering
\includegraphics[width=.45\textwidth]{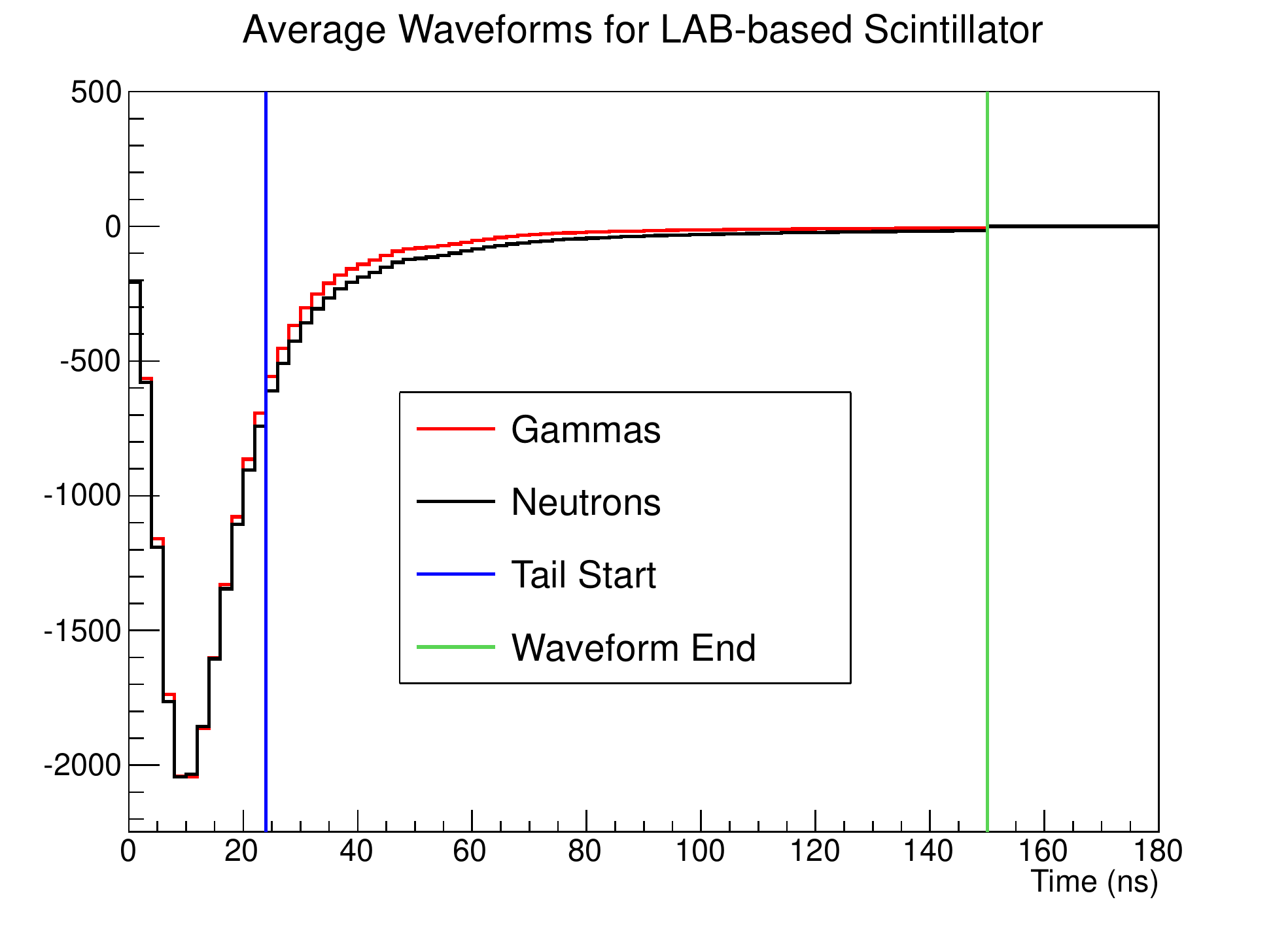}
\includegraphics[width=.45\textwidth]{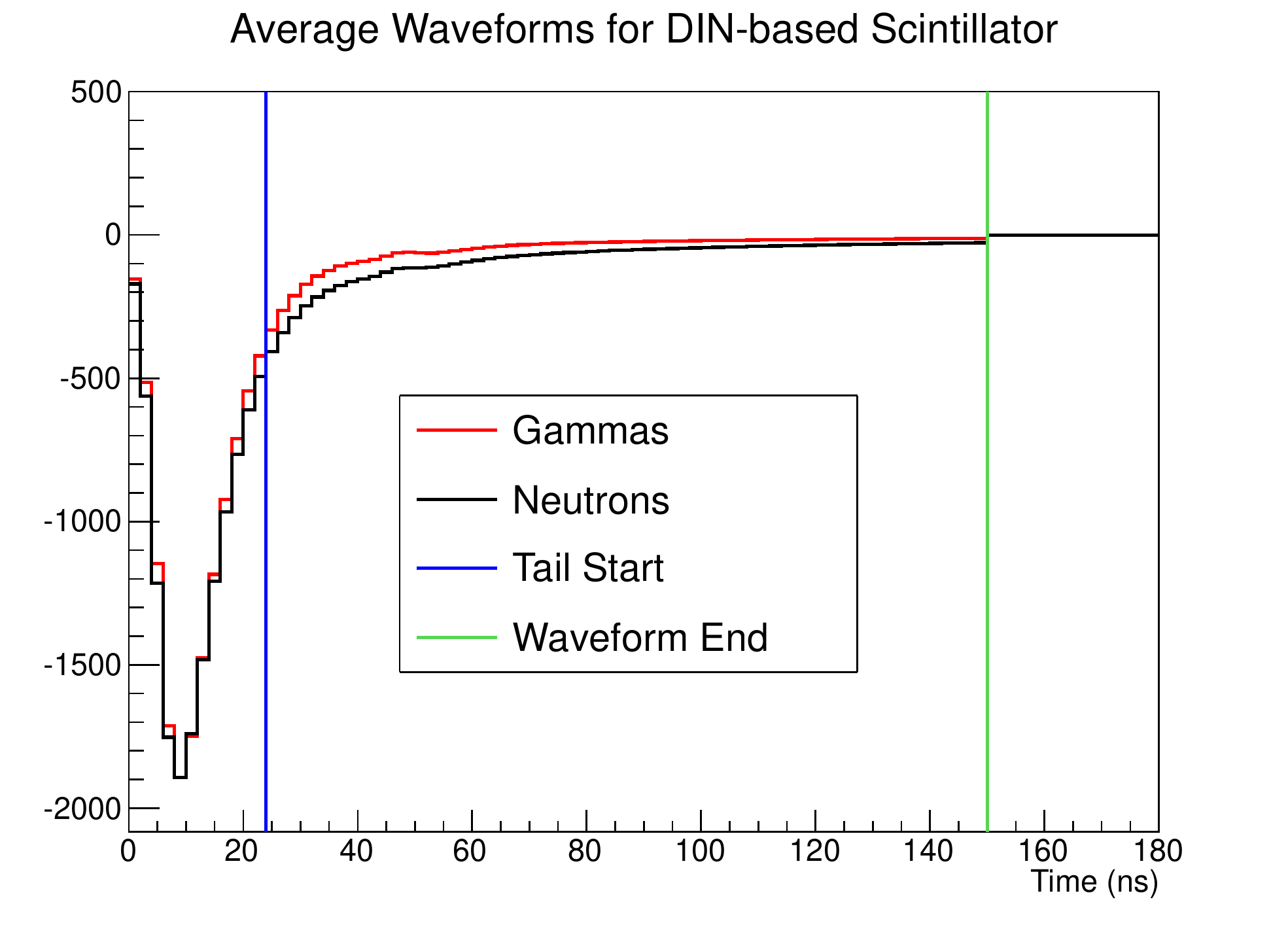}
\caption{\setlength{\baselineskip}{4mm} Average waveforms for neutrons (black) and gammas (red) in the LAB-based scintillator (left) and the DIN-based scintillator (right). The neutron waveforms show a longer tail in both scintillator cocktails. We define the tail as the portion of the waveform after the blue line.}
\label{AvgWaveforms}
\end{figure}

We can exploit the fact that neutron and gamma waveforms have different shapes (see average waveforms in Fig.~\ref{AvgWaveforms}) to define a pulse shape discrimination variable which is simply the amount of charge in the tail divided by the total charge in the whole waveform, often denoted Tail Q / Total Q. Neutrons have long tails and thus will generally have higher values of Tail Q / Total Q than gamma events. This fact will allow us to distinguish between the two types of events and reject fast neutrons. 

One prominent background to this measurement is misclassified gammas that are counted as neutrons due to accidental coincidence between two consecutive Cf decays. We already subtracted the effects using offline information to estimate the
rejection factor of the neutron events and the detection efficiency of the
gamma events.

\begin{figure}[h]
\centering
\includegraphics[width=.45\textwidth]{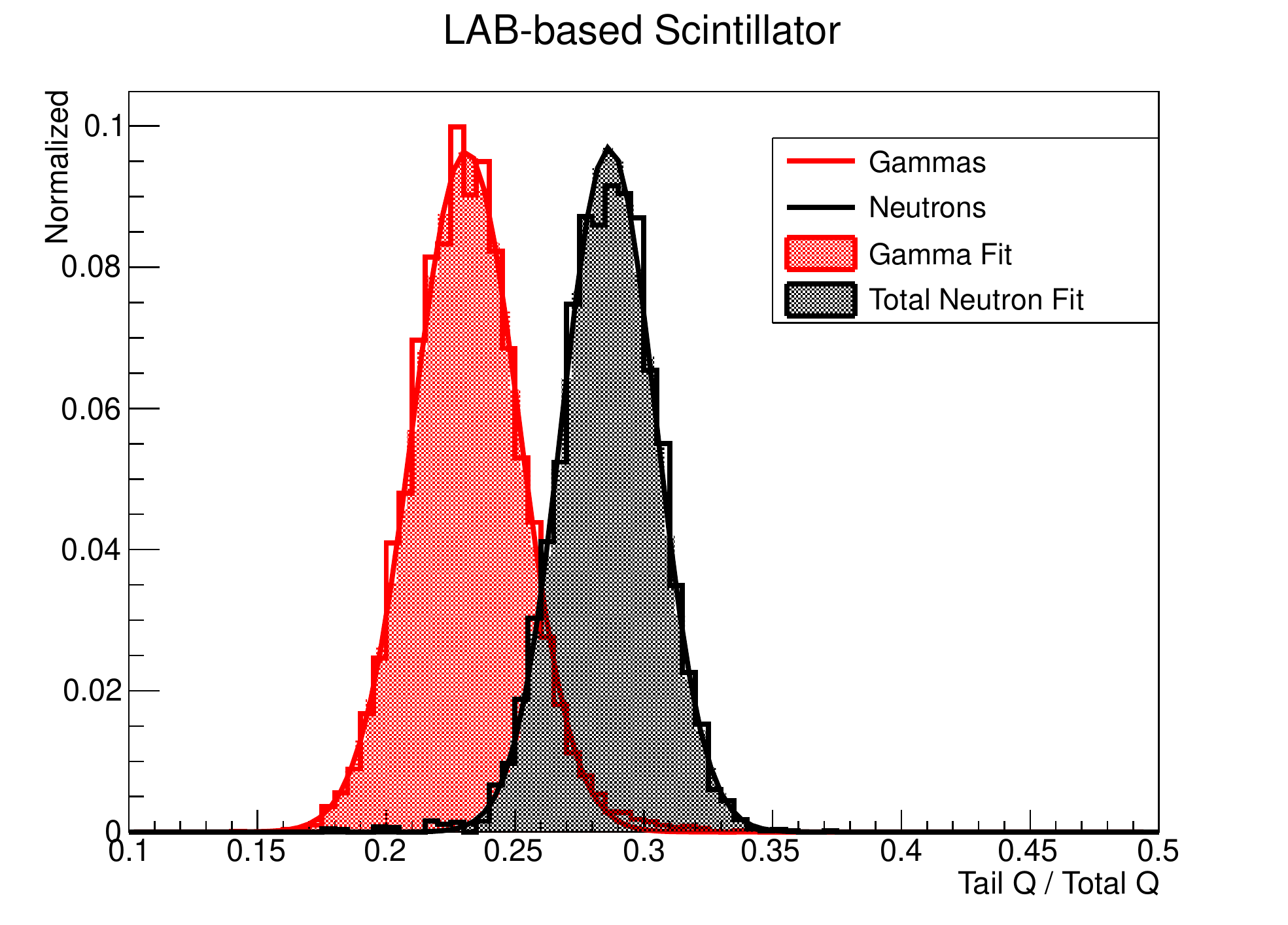}
\includegraphics[width=.45\textwidth]{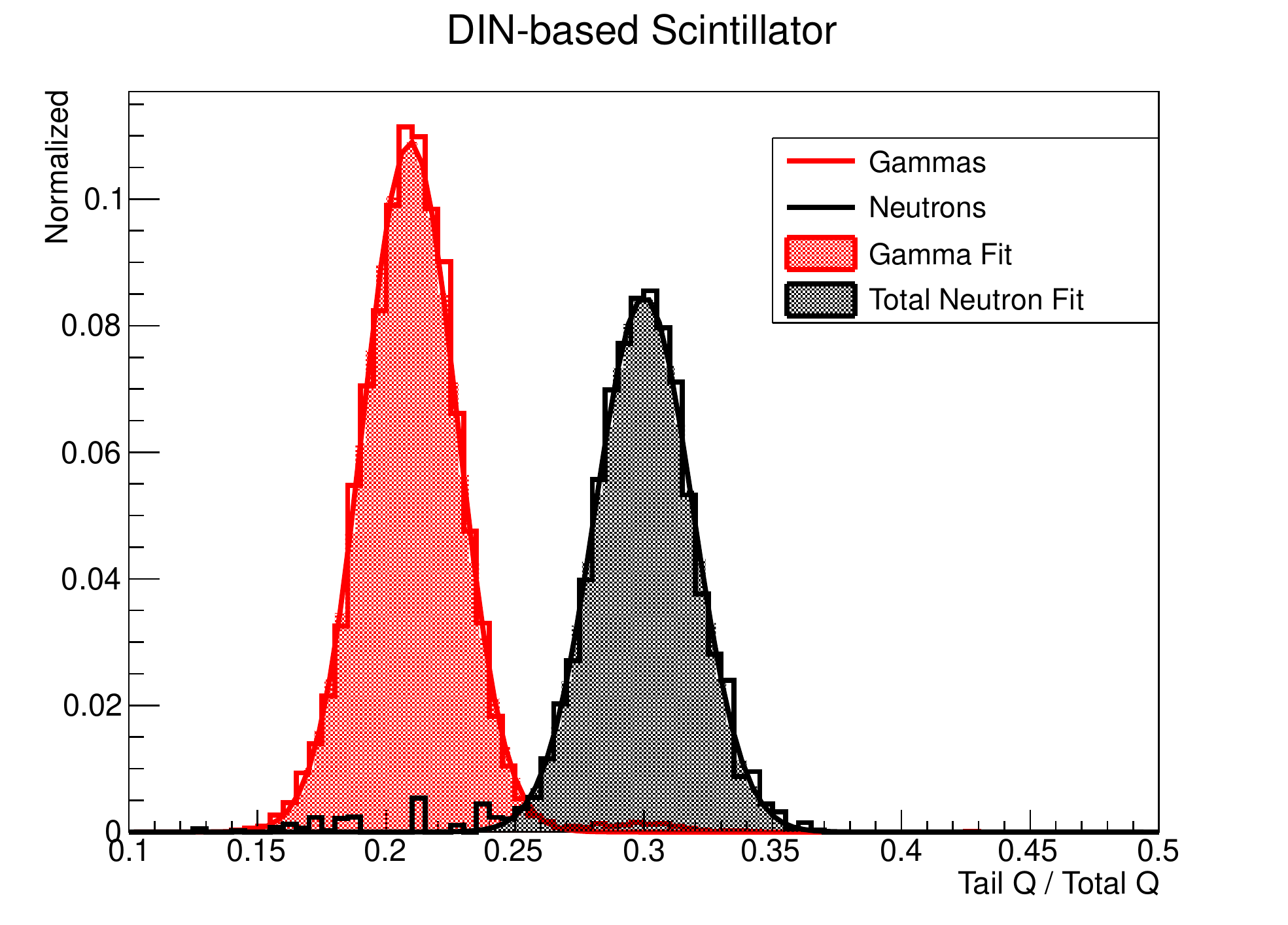}
\caption{\setlength{\baselineskip}{4mm} Measured distributions of Tail Q/Total Q for neutron (black histogram) and gamma events (red histogram) in the LAB-based scintillator (left) and the DIN-based scintillator (right). Misclassified gammas in the neutron population have been removed. Fits to the neutron and gamma distributions are shown as the filled Gaussians.}
\label{RatioQBackgroundRemoved}
\end{figure}

In Fig.~\ref{RatioQBackgroundRemoved}, the gamma and neutron distributions are shown side by side with the misclassified gamma events removed. It is this background-subtracted data that will be used to quantify the neutron rejection factor and gamma detection efficiency.

For this analysis, only events with between 200 and 800 photoelectrons (PEs) were used to assess the pulse shape discrimination performance of the scintillator cocktails. This population was chosen to include the highest energy neutron events ($\sim$ 10 MeV neutrons), but also to get sufficient statistics for background measurements. 

Using these events, we were able to achieve a neutron rejection factor of 100 with a gamma detection efficiency of 73.1\% for the LAB-based scintillator and 99.5\% with the DIN-based scintillator. With high energy events and better photocoverage in the full detector, these values will improve. We can also pair this pulse shape discrimination technique with Cerenkov light timing to get even better neutron rejection factors.

\subsubsection{\setlength{\baselineskip}{4mm} Properties of Diluted Liquid Scintillator, PSD + Cherenkov}
\indent

At this PAC meeting, the concrete recipe to make the liquid
scintillator is proposed to combine PSD and
Cherenkov method in order to have strong rejection power of the neutron events
induced by cosmic rays. The recipe is to use LAB (Linear alkylbenzene) +
0.5g/L PPO (secondary emission material).

\begin{figure}
 \centering
 \includegraphics[width=0.6\textwidth]{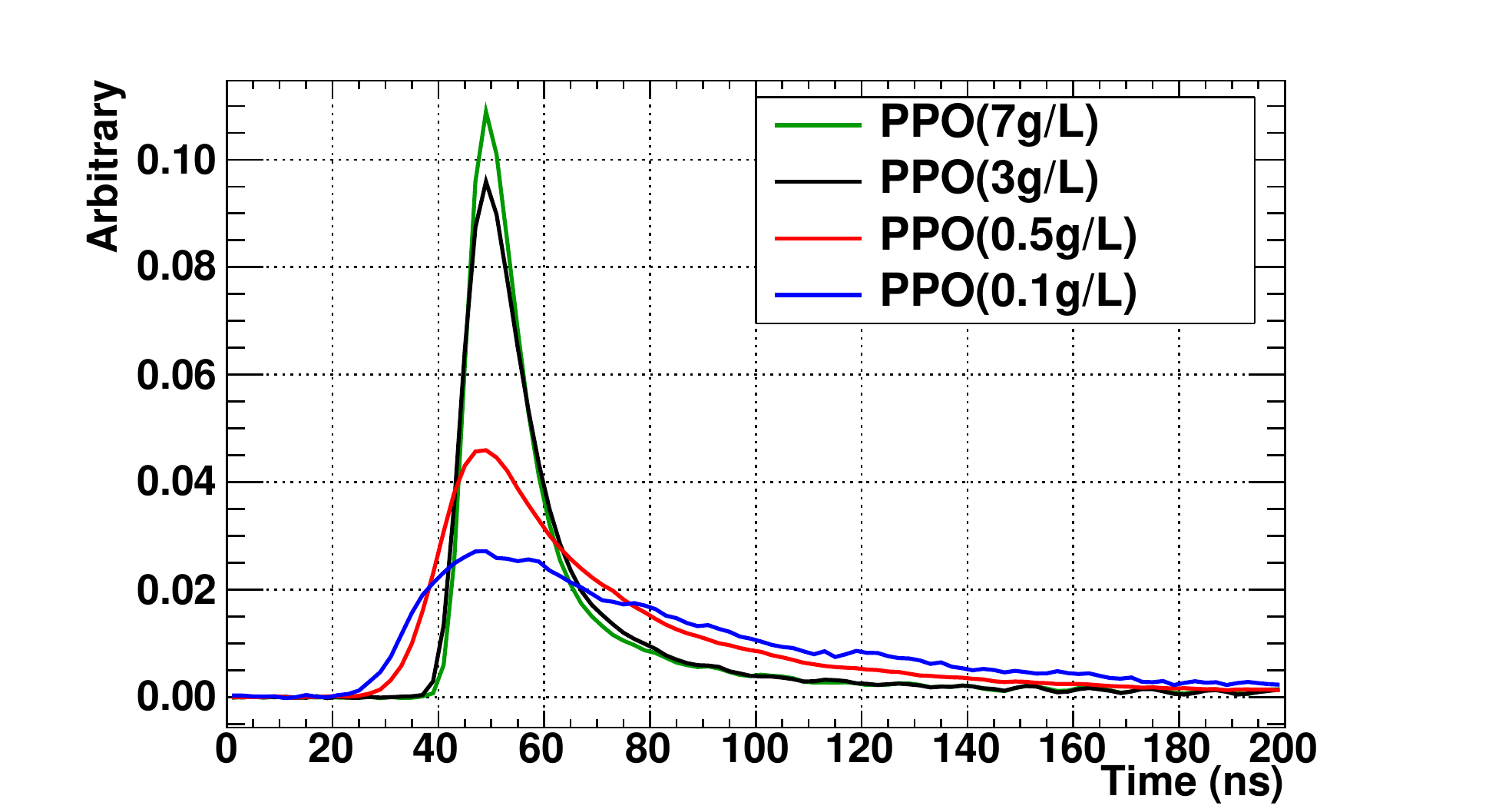}
 \caption{\setlength{\baselineskip}{4mm}
   PPO concentration dependence of the mean waveform. It is clear that
   the concentration affects not only light yield but also the light
   emission constant. 
 }
 \label{fig:PPOdep}
\end{figure}

Dr. Furuta (Tohoku University) found that the emission time constant
of the scintillation light depends on the concentration (density) of the
secondary light emission
materials such as PPO or b-PBD. Using different
concentrations of this material provides not only different scintillation light yields
but also different emission time constants.
Figure~\ref{fig:PPOdep} shows the dependence of the PPO concentration on the
light yield and the emission time constant. The emission time of scintillation
light is getting slower as the concentration is decreased. This means that
lower PPO concentration gives better condition for the Cherenkov method, while
PSD capability is better in the higher concentration condition. 

\begin{figure}
 \centering
 \includegraphics[width=0.6\textwidth]{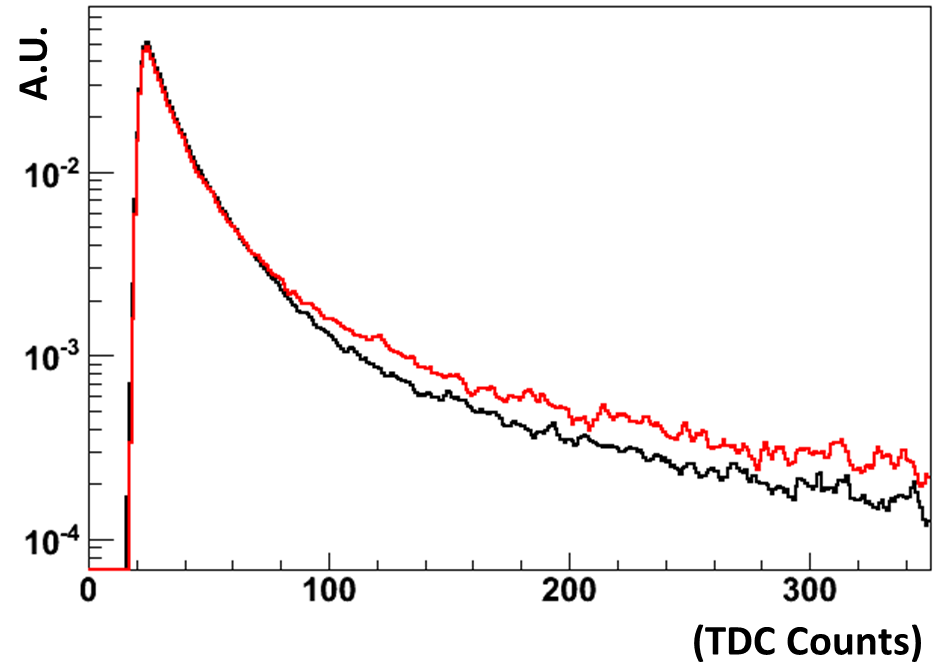}
 \caption{\setlength{\baselineskip}{4mm}
   The pulse shape difference between gamma events and neutron events
   measured by a vial size detector using a Cf radioactive source. The
   black line shows the gamma events, while the red line shows the neutron
   events. The horizontal axis corresponds to the TDC counts, which provides
   2ns / count. Later than the 80 counts ($\sim$160 ns), there are remarkable differences in pulse shape.
 }
 \label{fig:PSD_dil2}
\end{figure}

The pulse shape difference between gamma ray events and neutron events
measured by a vial size detector with Cf source is shown in
Fig.~\ref{fig:PSD_dil2}.
The black line shows the gamma events, while the red line shows the neutron
events. The horizontal axis corresponds to the TDC counts, corresponding to
2ns / count. After 80 counts ($\sim$160 ns), there are remarkable
pulse shape differences.
The difference is smaller than a normal PPO 3.0 g/L case, but the difference
still exists. The preliminary study for the PSD method using likelihood
gives good identifications between gamma and neutron events even in this
condition.

\begin{figure}
 \centering
 \includegraphics[width=0.7\textwidth]{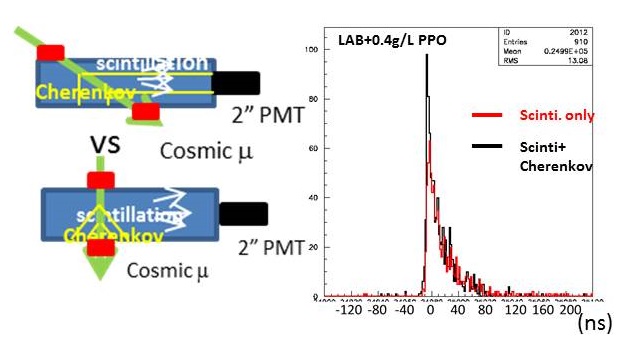}
 \caption{\setlength{\baselineskip}{4mm}
   The Cherenkov light yield measurement using the KEK
   test-stand~\cite{CITE:20PAC}. Using the two different setups shown in the
   left, the scintillation light timing and the scintillation timing
   + Cherenkov light timing was observed. Even in this condition, we
   can see the Cherenkov light in the fastest timing bin.
 }
 \label{fig:Cheren}
\end{figure}

The Cherenkov light yield compared to the scintillation light yield is measured by
KEK teststand~\cite{CITE:20PAC}. The analysis method is identical to those
used in the reference. All photons coming to the 2'' PMT are at one photo-electron
level because the mean light yield is about 0.4 p.e.
Figure~\ref{fig:Cheren} shows the results. The horizontal axis shows the
relative light emission timing with respect to the muon passing timing.
We still see the Cherenkov component in the fastest timing bin, but the
light yield ratio between the Cherenkov component vs. scintillation light is
almost one even in the fastest timing bin.

Using these results, we can estimate the rejection power of fast
neutron events using a simulation of the real JSNS$^2$ detector.
Note that this result set the PMTs apart from the light source
by $\sim$70cm. If the vertices are farther than 70 cm, the situation is
improved because the number of scintillation photons is
reduced by 1$/r^2$, where r is the distance between light source and a PMT,
while the number of Cherenkov photons are reduced by 1$/r$ due to the ring
image. For example, the ratio of Cherenkov and scintillation light
in the fastest timing bin is more than 10 times better in the case that
the vertex is located at the center of the detector (r$\sim$280cm).

%************************
\subsection{Veto System Design}
\indent

\begin{figure}[hbtp]
	\begin{center}
		\includegraphics[scale=0.35]{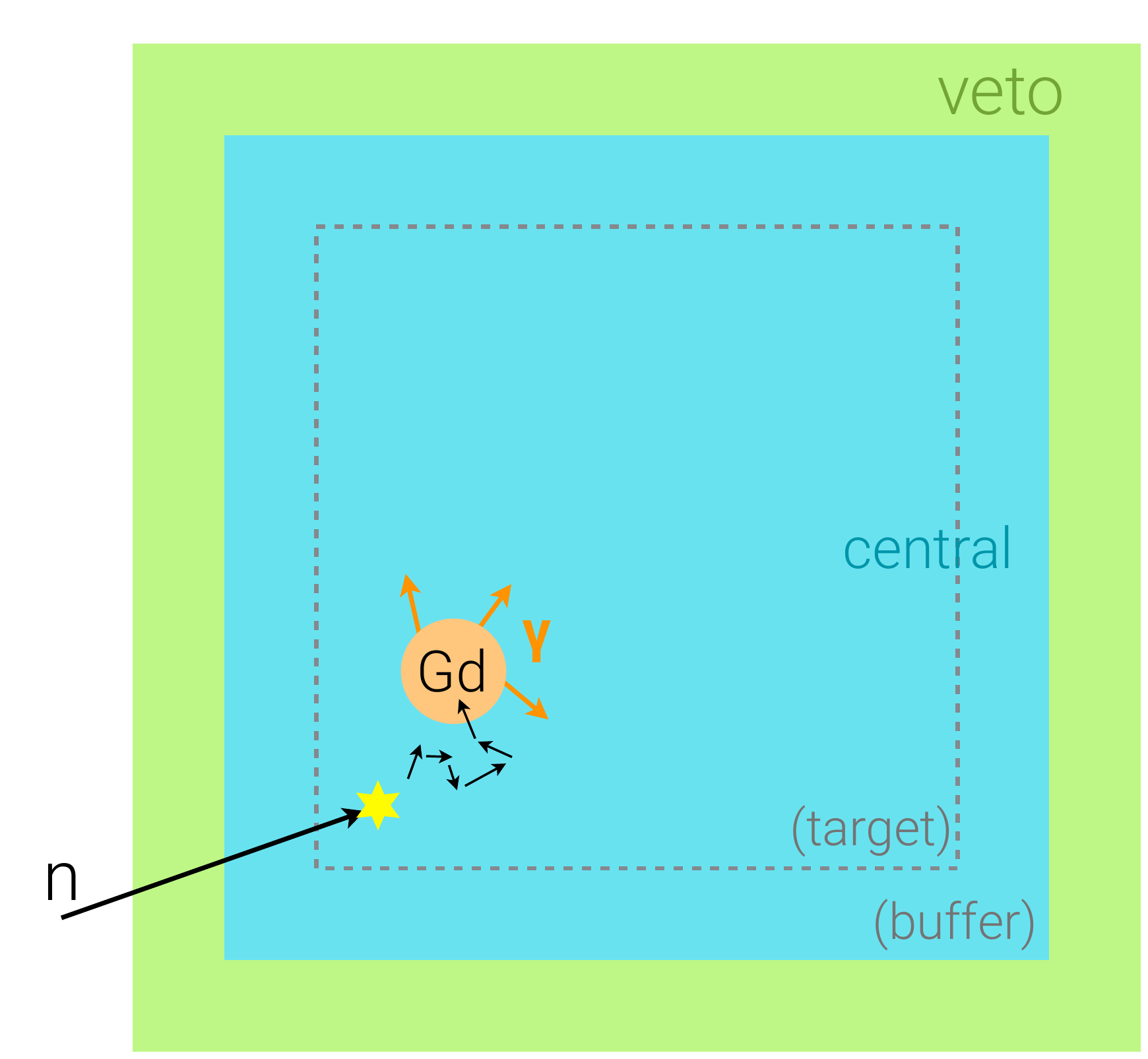}
		\caption{\setlength{\baselineskip}{4mm} A schematic view of the $\rm JSNS^2$ detector, and the delayed background induced by a fast neutron.}
		\label{fig_detector}
	\end{center}
\end{figure}

As shown in Fig.~\ref{fig_detector}, in the current detector design,
the fiducial target region filled with Gd-loaded liquid scintillator and
the buffer liquid scintillator layer are surrounded with an additional liquid scintillator layer to veto mainly charged particle from the outside.

In the previous status report \cite{CITE:SR_14NOV} for the 19th J-PARC PAC,
the veto layer was used to suppress the delayed backgrounds coming from
beam neutrons coming on the proton bunch timing.
Some of the fast neutrons on the bunch timing are thermalized in the detector
and captured by Gd to make delayed backgrounds (Fig.~\ref{fig_detector}).
Most of them produce some activity in coincidence with the bunch timing.
The veto system helps to reject these backgrounds.
In the previous study, the energy threshold to detect such activities
was set to 0.5 MeV for both the central region and the veto layer.

An energy threshold of 0.5 MeV is hard to utilize for the JSNS$^2$ detector
because of the energy range of the
environmental gammas (up to 2.6 MeV) and those of gammas from PMT glass,
the aclyric vessel, and liquid scintillator.
We thus study the case with a detector energy threshold of 3.0 MeV.

\begin{figure}[hbtp]
	\begin{center}
	  \includegraphics[scale=0.5]{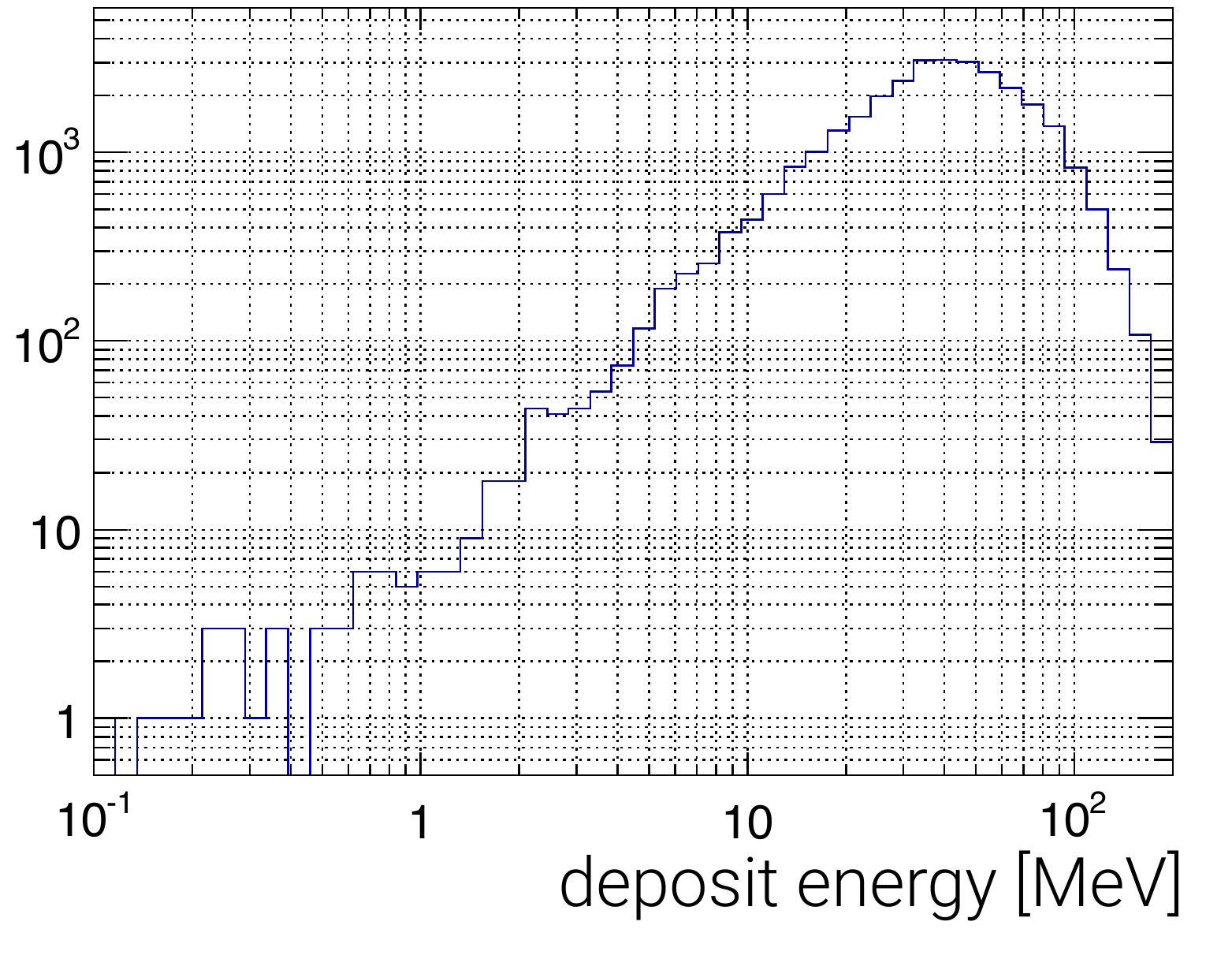}
		\caption{\setlength{\baselineskip}{4mm} The true energy deposit in the central region with beam fast neutrons coming on the bunch timing, which later make delayed background, obtained with a MC simulation. Note that this does not include quenching factors inside the liquid scintillator.}
		\label{fig_onbunchE}
		\includegraphics[scale=0.5]{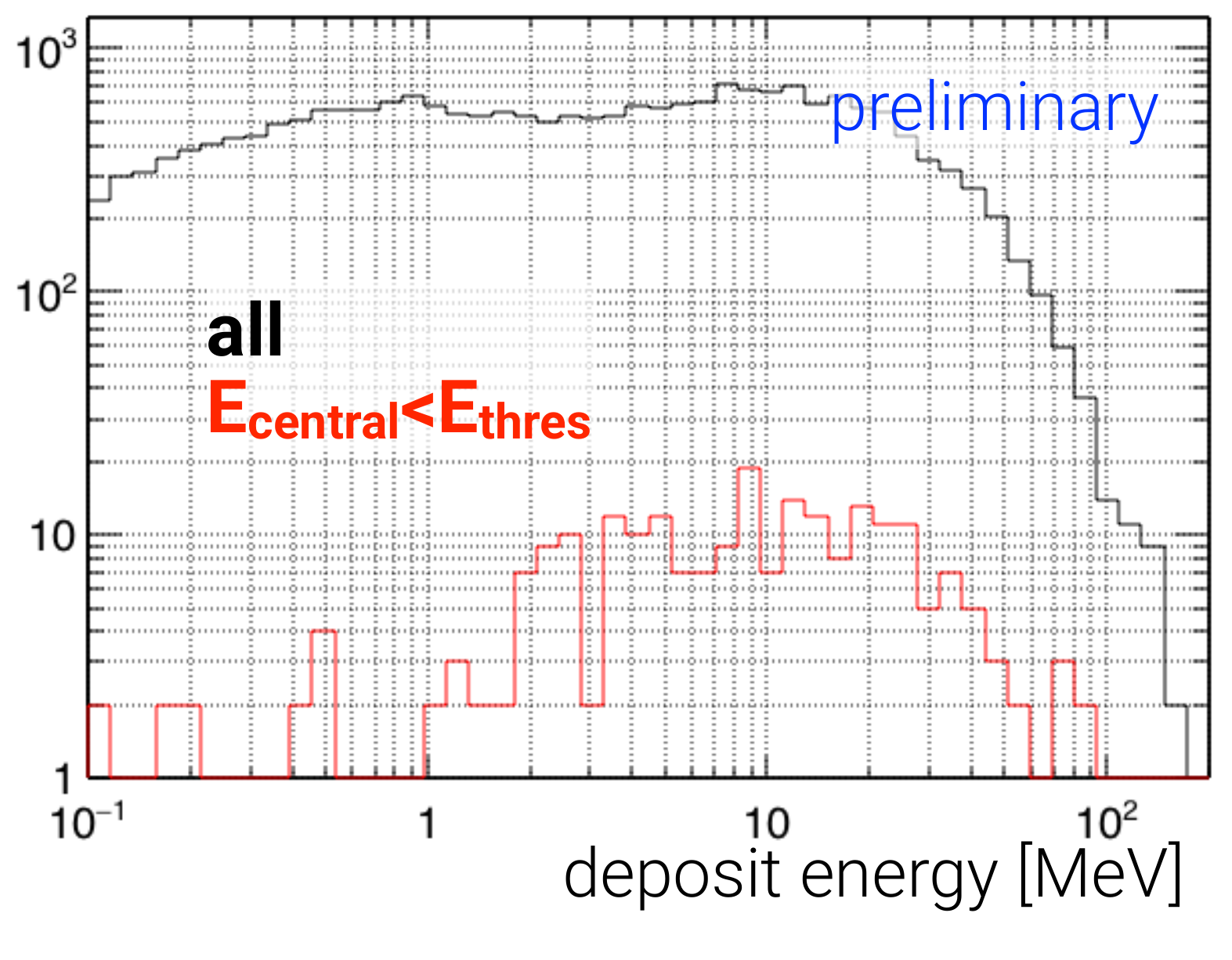}
		\caption{\setlength{\baselineskip}{4mm} The energy deposit in the veto layer with beam fast neutrons coming on the bunch timing, which later make delayed backgrounds, obtained with a MC simulation. Black corresponds to 
                  all the activity, while the red shows the remaining background
		  after the central energy cut. The energy threshold for the central region is 3 MeV. Note that this does not include quenching factors inside the liquid scintillator.}
		\label{fig_onbunchE_veto}
	\end{center}
\end{figure}

Figure~\ref{fig_onbunchE} shows the energy deposit in the central region of the detector
with the beam fast neutrons coming in coincidence with the proton bunch timing, which later make
delayed background, obtained with a MC simulation. Note that this energy
deposit does not include the quenching factor in the liquid scintillator,
thus we will include it in the next publication.
The assumed energy distribution and the flux of the beam fast neutrons are
based on the measurement at the candidate detector location, and it is same as the
previous study (described in the article~\cite{CITE:SR_14NOV}).
In terms of the energy deposit distribution of the central region,
there is no need to apply the energy threshold of 0.5 MeV, therefore
3.0 MeV is a reasonable threshold. 

The energy deposit distribution in the veto layer in coincidence with the bunch timing
for all of (black) and the remaining background after the central energy cut
with 3.0 MeV (red) is shown in Fig.~\ref{fig_onbunchE_veto}.
An energy threshold of 3.0 MeV
is good enough to keep a good rejection factor of the remaining background.

Although the quenching factor should be included to obtain the final number,
more than 99$\%$ of the delayed background created from beam neutrons 
can be tagged with a 3.0 MeV energy threshold.

\subsubsection{Hardware Design Study}

\begin{figure}[hbtp]
	\begin{center}
		\includegraphics[scale=0.5]{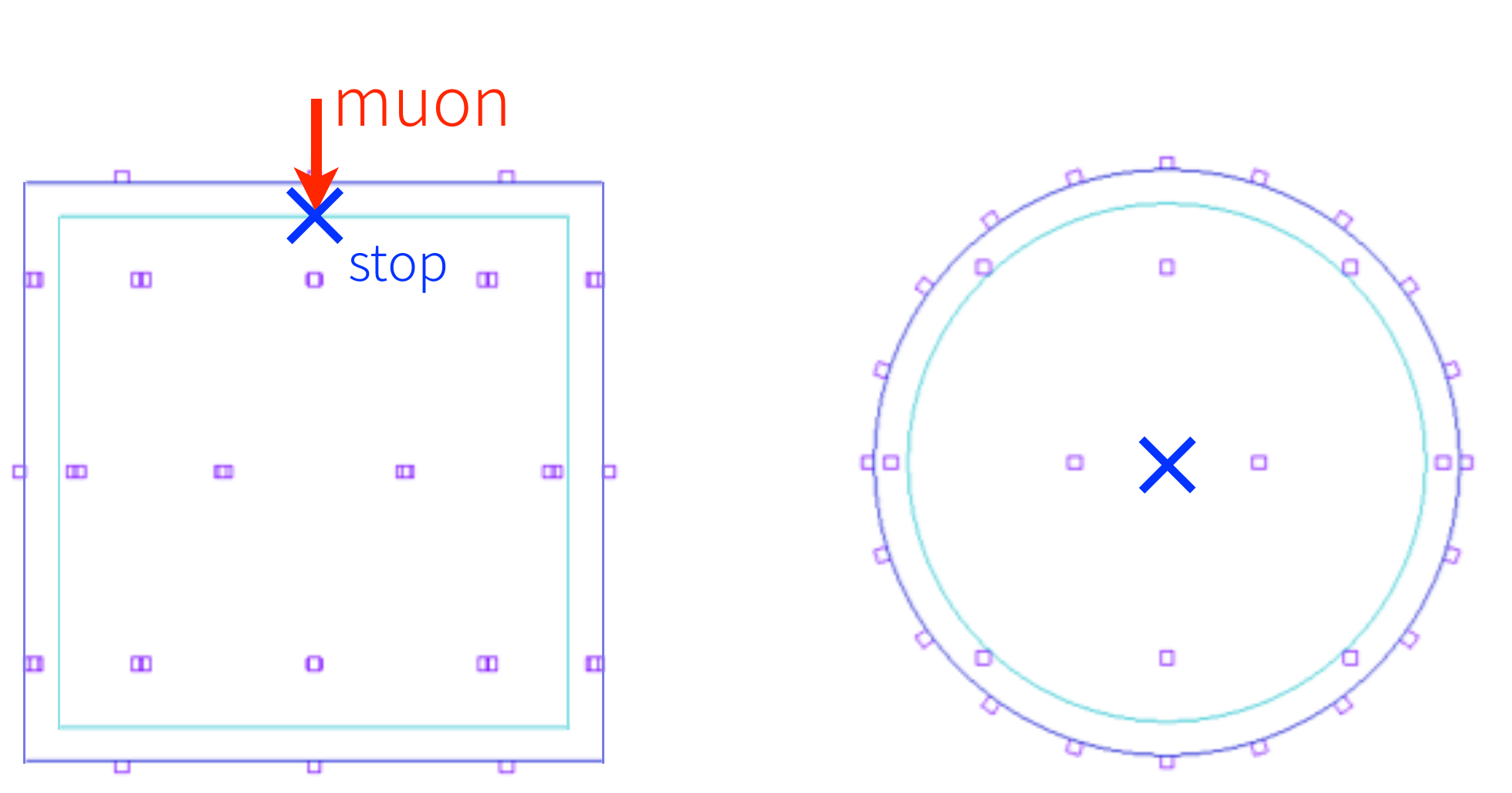}
		\caption{\setlength{\baselineskip}{4mm}
                 The MC setup for the veto performance evaluation.
		The inner and outer surfaces of the veto layer are covered by reflecting sheets and the layer is viewed by 50 5" PMTs.}
		\label{fig_setup}
		\includegraphics[scale=0.5]{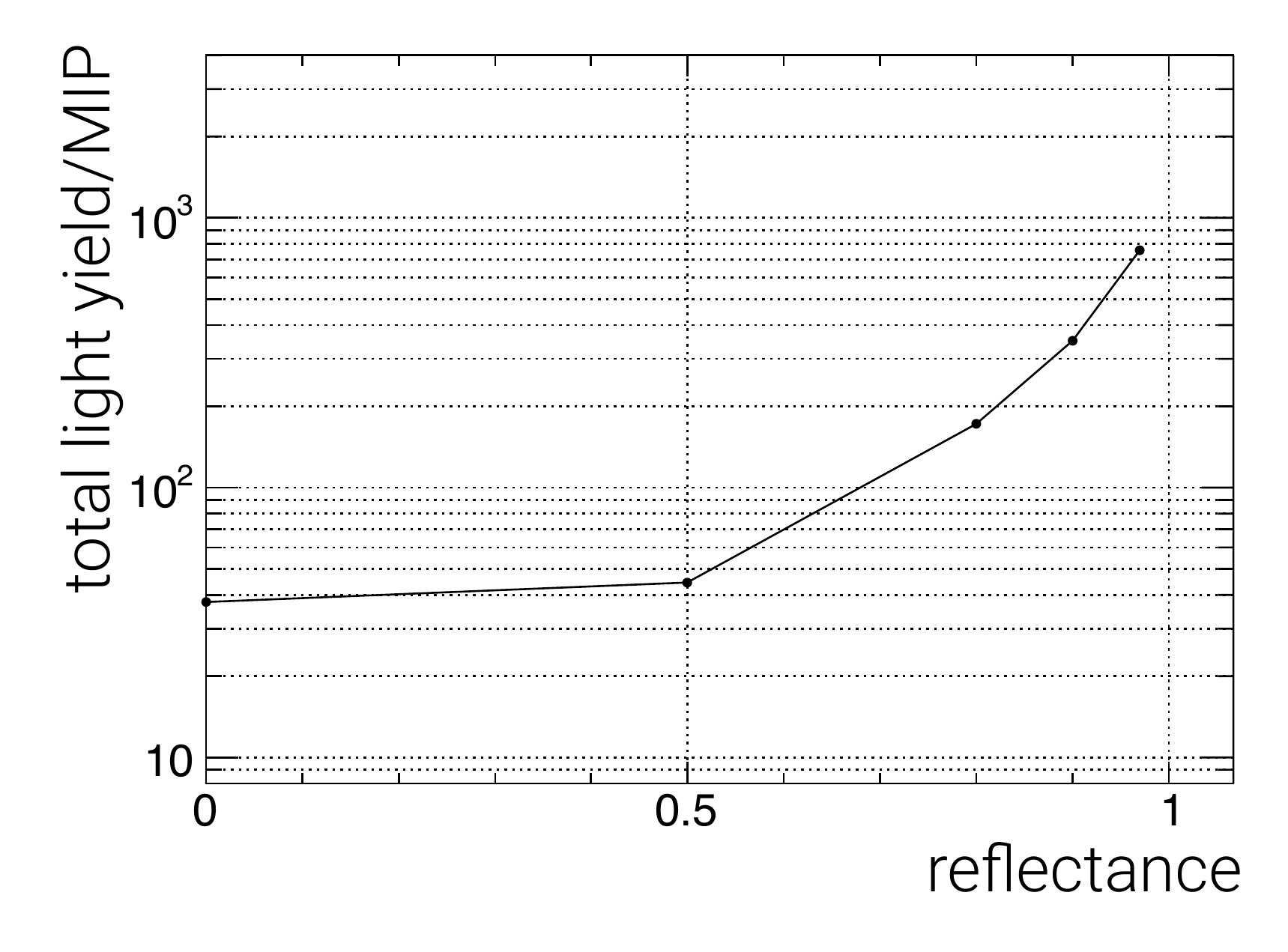}
		\caption{\setlength{\baselineskip}{4mm}
                The expected total light yield for MIP particles as a function of the reflectance at the surface of the veto layer.}
		\label{fig_lightYield}
	\end{center}
\end{figure}

The veto performance has been studied with a MC simulation.
Figure~\ref{fig_setup} shows the tested setup.
The veto layer (25 cm thick) surrounds the central region.
The scintillation light from the veto layer is viewed by 50 5" PMTs (top and bottom: 10 each, side: 30).
The inner and outer surfaces of the veto are covered by reflecting sheets.
One candidate of the reflection sheet is called VM2000, which gives more than 97\% reflectance for $>\!400$ nm (98.3\%@430nm)\cite{CITE:VM2000}.
We evaluated the expected light yield as a function of the reflectance of the reflection sheet.
Figure~\ref{fig_lightYield} shows the result.
The expected light yield is about 350 photoelectrons\footnote{Assuming 10000 photon/MeV for the liquid scintillator, and the relevant QE for the PMT} for MIP particles (energy deposit of 40 MeV) with a 90\% reflecting sheet; it is about 10 times of without the reflecting sheet.

\begin{figure}[hbtp]
	\begin{center}
		\includegraphics[scale=0.7]{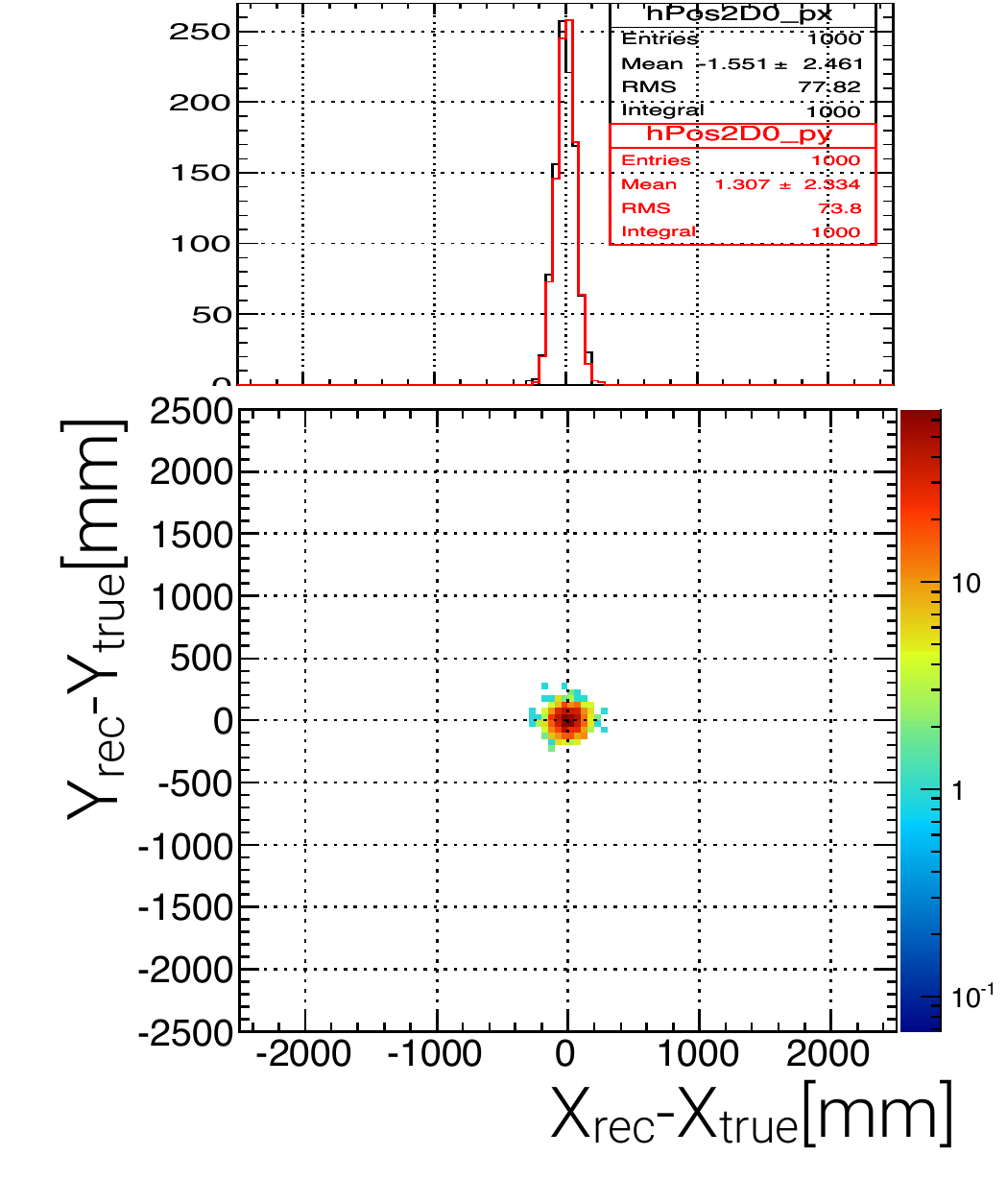}
		\caption{\setlength{\baselineskip}{4mm}
                  The residual distribution of the reconstructed positions.}
		\label{fig_posres}
	\end{center}
\end{figure}

By using a high-reflection sheet to increase light-yield, light distribution was spread out over the whole veto layer,
and some of the scintillation light can be detected even on the opposite side to the incident plane.
The position resolution was thus also studied.
With the configuration described above, the position resolution is about 7.5 cm for MIP particles as shown in Fig.~\ref{fig_posres}.

%************************

\subsubsection{An Alternative Option for the Veto}
\indent

As mentioned in the status report in 2015~\cite{CITE:20PAC}, we have an option
to use SiPMs in the veto region. Here we assume to segment the veto layer with
30cm$\times$30cm in order to have strong spatial resolution ($\sim$ 700 SiPMs are
needed in this case).
A preliminary study to use SiPMs (12mm$\times$12mm) and a reflector with 90$\%$ reflectivity in all
segmented area provides an acceptance of scintillation photons at 18 photons / MeV. 
This acceptance is good enough to detect the particles with a MeV.

\subsection{Software / Simulation}
%**********************
\subsubsection{The Reactor Analysis Tool (RAT)}

The Reactor Analysis Tool (RAT) is a simulation framework which adds new physics lists to Geant4 \cite{RATDocs}. In particular, RAT adds many features which allow for the detailed simulation of scintillation light. RAT makes it possible to input many of the most important parameters that characterize a scintillator including absorption length, refractive index, scintillation spectrum, and many more. With the ability to naturally specify all of these parameters, it is possible to run more accurate simulations to characterize a detector and all relevant backgrounds.

\subsubsection{Reproducing Data from the KEK Test Stand}

It has been shown in previous status reports that the background rates in JSNS$^2$ are manageable if a neutron rejection factor of 100 can be achieved \cite{CITE:SR_14NOV, CITE:20PAC}. Recoil protons in the 20-60 MeV range created by fast neutrons will not produce Cerenkov light and thus can be rejected using Cerenkov light timing. This technique was used successfully by LSND and needs to be independently verified for the scintillator under consideration for JSNS$^2$.

\begin{figure}[h]
\centering
\includegraphics[width=\textwidth]{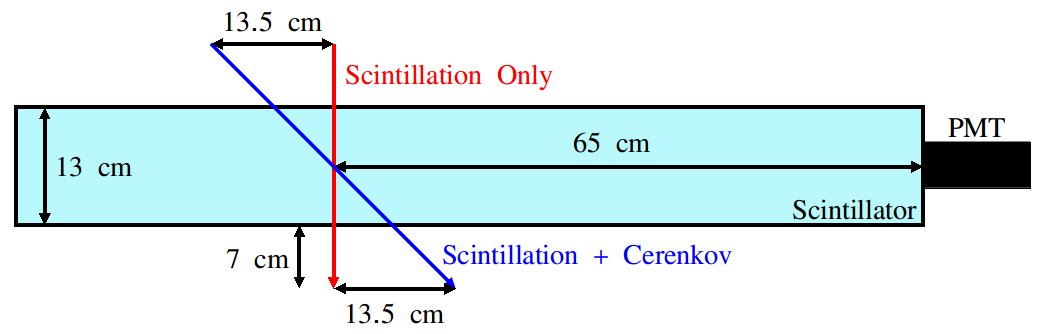}
\caption{\setlength{\baselineskip}{4mm} A schematic diagram of the KEK test stand geometry used in RAT. The red path represents 4 GeV muons in the scintillation only configuration and the blue path represents 4 GeV muons in the scintillation + Cerenkov configuration.}
\label{KEKGeo}
\end{figure}

Measurements were done at KEK to quantify the fractions of Cerenkov and scintillation light in a dilute liquid scintillator cocktail. Scintillator paddles were used to select cosmic rays passing through a tube of liquid scintillator (1 m in length, 13 cm diameter) in two different configurations. These two configurations are shown in Fig.~\ref{KEKGeo}. In the scintillation only configuration, muons pass straight through the detector so that no Cerenkov light reaches the PMT at the end of the scintillator tube. In the scintillation + Cerenkov configuration, the muon tracks enter the scintillator at an angle so that some of the Cerenkov light will reach the PMT. More details can be found in previous status reports which contain the test results~\cite{CITE:20PAC, CITE:21PAC}.

\begin{figure}[h]
\centering
\includegraphics[width=.65\textwidth]{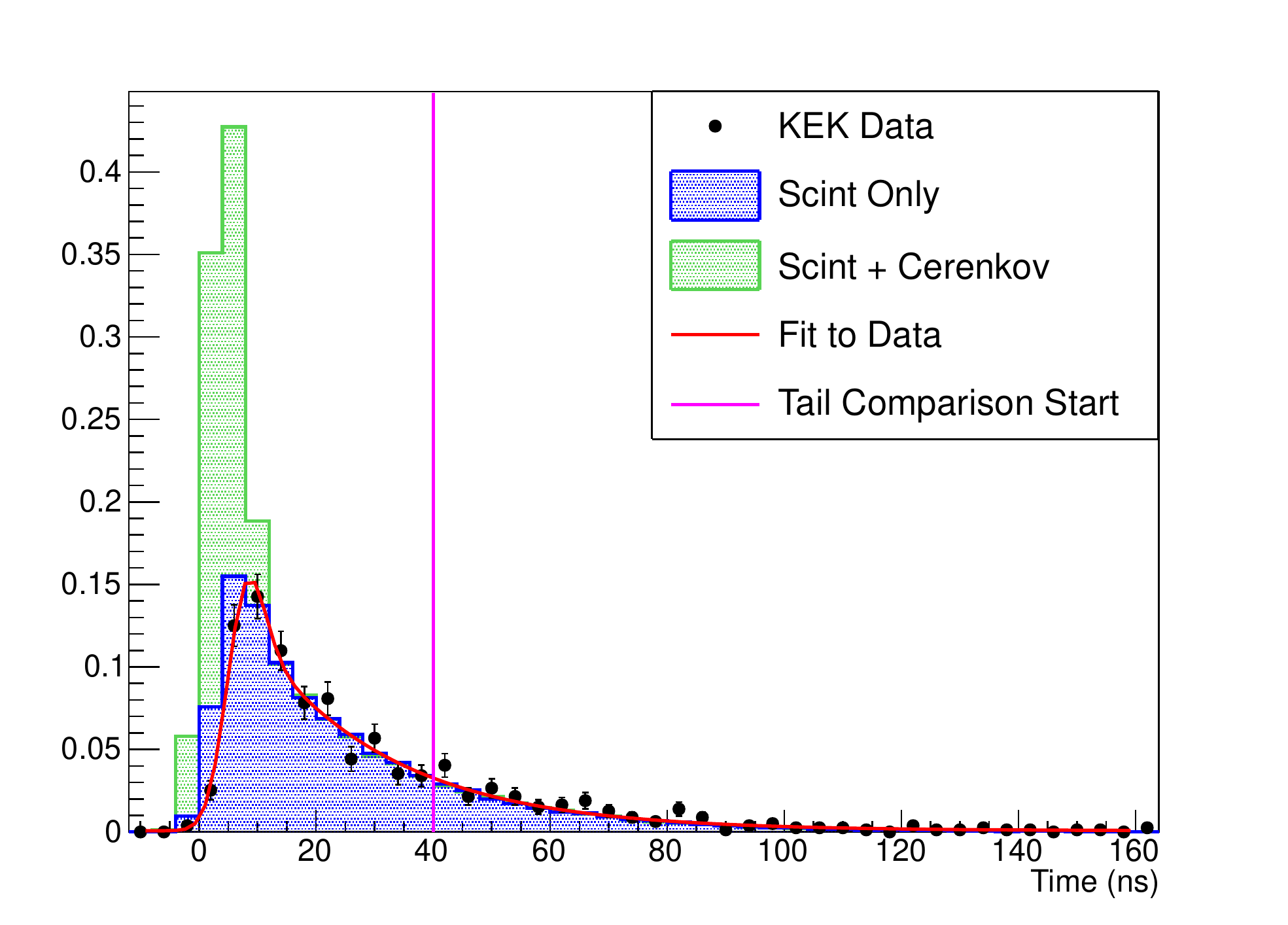}
\caption{\setlength{\baselineskip}{4mm} A comparison between RAT simulation results and KEK data in the scintillation only configuration (black points). The blue filled histogram is simulation data in the scintillation only configuration and the green filled histogram is data from the scintillation + Cerenkov configuration. The red line is a fit used to extract the time constants from the data. Bins to the right of the pink line were used to normalize the scintillation contributions to the simulation results.}
\label{DataSimComparison}
\end{figure}

We are interested in replicating the measurements from the KEK test stand using RAT to verify that the simulation is working correctly and to ensure that we can reproduce the scintillator behavior in simulation. The geometry shown in Fig.~\ref{KEKGeo} was constructed in RAT and 4 GeV muons were simulated through the two paths shown. The scintillation time constants used in the simulation were taken from fits done to measurements made at the KEK test stand in the scintillation only configuration. This fit is shown as the red line in Fig.~\ref{DataSimComparison}. Since the scintillator used in the KEK test stand was dilute (0.5 g/L PPO as opposed to 3.0 g/L PPO for DBLS), the light yield was changed from 10,000 photons/MeV to 4,000 photons/MeV.

In order to control for the fact that there is more scintillation light in the scintillation + Cerenkov configuration than in the scintillation only configuration due to the longer path length, the waveform tails (where there is no Cerenkov light) were used as a ``side-band" to set the normalization. All bins to the right of the pink line in Fig.~\ref{DataSimComparison} were integrated and the two histograms were scaled so that these integrals would be equal.

The scintillation only data produced by RAT matches the scintillation only data taken at KEK. There is also a clear excess of Cerenkov light at early times that could be used to distinguish between different types of particles.
This tendency is clearly similar to the Fig.~\ref{fig:Cheren} although more quantitative comparison will be performed. 
The ratio of Cerenkov to scintillation light at early times is highly dependent on the particle energy so a more realistic cosmic ray generator is being investigated to improve these results.

%**********************

\section{MLF 2015AU0001 Test-Experiment}
\indent

This test-experiment has been done from May to June in 2016 at the MLF
3rd floor\footnote{\setlength{\baselineskip}{4mm} Collaborators of this test experiment are followings; S.Meigo, S.Hasegawa (JAEA), E.Iwai, T.Maruyama (KEK), T.Hiraiwa, T.Shima (Osaka RCNP), H.Furuta, Y.Hino, F.Suekane (Tohoku), J.Spitz (U of Michigan).}.

The goal of the experiment is to measure the PID of the background events, which
occur in coincidence with the proton bunch timing, using 1.6 L of liquid scintillator. This
liquid scintillator is good at separating neutrons from gamma events with
PSD. We assumed all of this activities are coming
from neutrons[2],
but we may have a smaller number of delayed neutron background and a
smaller neutrino oscillation signal detection efficiency in the case that there
are a large number of gamma events. Therefore, this measurement is crucial for
the real JSNS$^2$ experiment.

\begin{figure}
 \centering
 \includegraphics[width=0.7\textwidth]{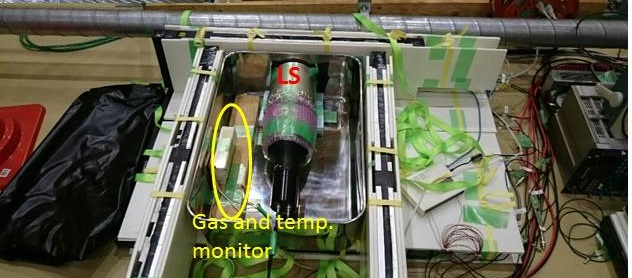}
 \caption{\setlength{\baselineskip}{4mm}
   The setup of the 1.6L test measurement.
 }
 \label{fig:1.6setup}
\end{figure}

Figure~\ref{fig:1.6setup} shows the setup of the experiment. This detector was
put in a location similar to that of the current best candidate JSNS$^{2}$ detector location.
The 1.6L LS detector was put inside a oil protection box made of stainless
steel, and the box was surrounded by the veto counters made of plastic
scintillator.
Inside the box, there were temperature and gas level monitors to detect the
liquid leak in the case of an emergency. This information was also monitored by those
who in the MLF control room, and warning alarms were ready to alert operators in case of a liquid leak. Notably, it is envisioned that this system can be used for the real experimental case.

\begin{figure}
 \centering
 \includegraphics[width=0.7\textwidth]{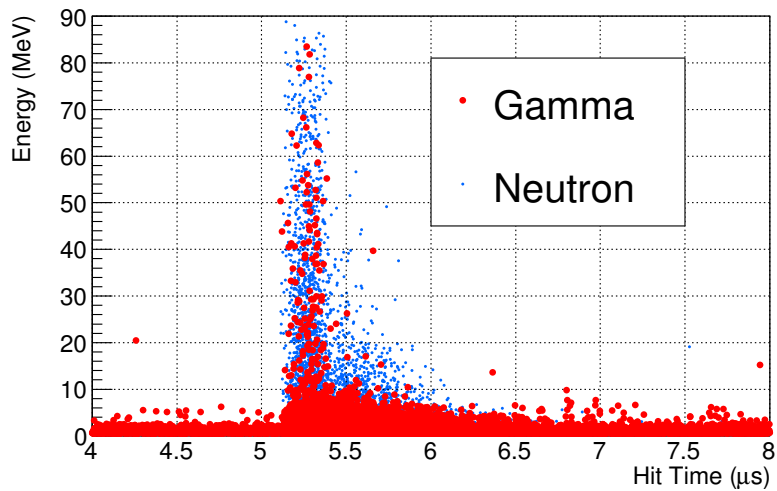}
 \caption{\setlength{\baselineskip}{4mm}
   A preliminary result from the 1.6 L test measurement at the MLF. Blue points show the
   neutron events without cosmic veto hits and red points show 
   gamma ray events. Horizontal axis shows the event timing, while the vertical axis
   corresponds to the energy of the events.
 }
 \label{fig:1.6result}
\end{figure}

Figure~\ref{fig:1.6result} shows the preliminary results of the 1.6 L experiment.
The horizontal axis shows the event timing, while the vertical axis corresponds
to the energy of the events. Blue points show the neutron events without cosmic
veto hits and red points show the hits of those of gamma rays.
As can be seen in Fig.~\ref{fig:1.6result},
gamma events are visible.
The gamma events are mainly apparent in the faster timing region, as 
compared to the neutron events, with respect to the proton bunch timing.
This is a good indication for the real JSNS$^2$ experiment because the number of
delayed background events induced by beam neutrons should be reduced,
and the time window of the prompt signal can be close to the proton
event timing.
Further quantitative statements will be shown in a publication in near future.

\section{Summary}
\indent

The most significant news for the JSNS$^2$ experiment is that the project received funding to construct the first of two 25 ton fiducial volume detector modules. We aim to start the JSNS$^2$ experiment in 2018-2019 with
one detector. Currently, a concrete timescale and the budget of each experimental component
is being estimated.

The studies for the TDR are also in good shape.
Realistic MC simulation studies for the PSD capability using Daya-Bay
type LS have been performed. In particular, it has been found that the effect due to noise is small if each channel is independent. 

The primary candidate for the recipe to use both PSD and Cherenkov technique
is LAB+0.5g/L PPO. The preliminary proto-type tests at Tohoku University
and KEK show that this recipe can provide both PSD and Cherenkov capabilities
at the same time. 
We will determine the final recipe to make the liquid
scintillator and present this in the upcoming TDR.

The veto design has been extensively discussed in this status report in order to demonstrate the ability to
reject neutron and cosmic ray background events in the full JSNS$^2$ detector.

For the optical simulation, the RAT framework has been studied.

The test experiment, MLF 2015AU0001, has been performed from May-20 2016.
This test aims to measure the identification of background particles arriving during or just after the MLF proton bunches.
We have preliminary results, and will publish the paper soon.

%%%%%%%%%%%%%%%%%%%%%%%%%%%%%%%%%%%%%%%%%%%%%%%%%%
% bibliography
%%%%%%%%%%%%%%%%%%%%%%%%%%%%%%%%%%%%%%%%%%%%%%%%%%

\end{document}